# The density maximum of He$^4$ at the lambda point modeled by the stochastic quantum hydrodynamic analogy

Piero Chiarelli

*National Council of Research of Italy, Area of Pisa, 56124 Pisa, Moruzzi 1, Italy*

*Interdepartmental Center "E.Piaggio" University of Pisa*
Phone: +39-050-315-2359
Fax: +39-050-315-2166

Email: pchiare@ifc.cnr.it.

**Abstract.** The lambda point in liquid He$^4$ is a well established phenomenon acknowledged as an example of Bose-Einstain condensation. This is generally accepted, but there are serious discrepancies between the theory and experimental results, namely the lower value of the transition temperature $T_\lambda$ and the negative value of $dT_\lambda/dP$. These discrepancies can be explained in term of the quantum stochastic hydrodynamic analogy (QSHA). The QSHA shows that at the He$^4_\mathrm{I} \rightarrow$ He$^4_\mathrm{II}$ superfluid transition the quantum coherence length $\lambda_c$ becomes of order of the distance up to which the wave function of a couple of He$^4$ atoms extends itself. In this case, the He$^4_2$ state is quantum and the quantum pseudo-potential brings a repulsive interaction that leads to the negative $dT_\lambda/dP$ behavior. This fact overcomes the difficulty to explain the phenomenon by introducing a Hamiltonian inter-atomic repulsive potential that would obstacle the gas-liquid transition.

# 1. Introduction

To explain the He$^4_\mathrm{I} \rightarrow$ He$^4_\mathrm{II}$ superfluid transition London [1] in 1938 made the hypothesis that He$^4$ lambda point might be an example of Bose-Einstain (BE) condensation (BEC). This hypothesis was based upon the similarity between the shape of the heat capacity of an ideal boson gas at the BEC transition and the data for the He$^4_\mathrm{I} \rightarrow$ He$^4_\mathrm{II}$ transition. This convincement was reinforced by the observation that there is no similar phase transition in the Fermi liquid He$^3$. However, even if the basic BEC hypothesis is acknowledged, looking in details some discrepancies exist [2]. Among those, two are the majors: (1) the calculated BE transition temperature $T_B$ for an ideal gas is 3.14 K while the measured one for the He$^4$ is of 2.17K. (2) The variation of the transition temperature $T_\lambda$ with pressure is negative and is opposite in sign to that expected from the BEC. The standard way out is to address the differences to the fact that the BEC theory is applied to an ideal gas while the He$^4$ is clearly not, since it shows a van der waals-like liquid gas phase transition. Therefore, the inter-molecular potential must be taken into account when we calculate the transition temperature $T_B$ and its variation with temperature.

The BEC theory [3] affirms that below the BE temperature $T_B$ the number of particles, $N_e$, in the excited state reads

$$N_e = h^{-3} \int \frac{1}{\exp[\varepsilon/kT]-1} d^3q\, d^3p \qquad (1)$$

and hence

$$N_e(T) = 2{,}612 \, V \frac{2\pi(2m)^{3/2}}{h^3} T^{3/2}. \tag{2}$$

At the BE temperature $T_B$, it is assumed that all the particles go in the excited state so that

$$T_B = \frac{h^2}{2\pi mk} \left(\frac{N}{2{,}612 V}\right)^{3/2}. \tag{3}$$

As equation (3) shows, the increase of pressure, leading to the volume V decrease, will bring to the increase of $T_B$. This contradicts what is experimentally observed at lambda point where $dT_\lambda/dP$ is negative.

By considering the van der Waals state equation

$$P = \{n k T/ (V - n b)\} - a n^2 /V^2, \tag{5}$$

(where n is the number of molecules, b is the fourfold atomic volume and a is the mean inter-molecular potential energy derived by using the rigid sphere approximation given by (21-22)) that for punctual particles (i.e., b = 0) with an attractive potential (i.e., a>0) reads

$$P = n k T/ V - a n^2 /V^2 \tag{6}$$

we can see that the pressure decrease $- a n^2 /V^2$ is a consequence of the attractive intermolecular potential. This is equivalent to a compression of the ideal gas and, since the integration in (1) is carried out on the system volume, we can say that a cohesive intermolecular potential reduces the system volume and by (3) that $T_B$ increases.

Therefore, given the ideal gas pressure $P_{IG} \cong n k T/ V$, the variation of BEC temperature $\Delta T_B$ has the same sign of the pressure variation $\Delta P$ (with respect to the real gas) according to the expression

$$\Delta T_B \propto \Delta P = P_{IG} - P \cong a n^2 /V^2 > 0 \tag{7}$$

Moreover, given $dT_B/dP \sim d\Delta T_B/dP$ it follows that

$$dT_B/dP \propto a n^2 \, d(V^{-2})/dP > 0 \tag{8}$$

since V decreases with the pressure.

Feyman [4] in 1953 and later Butler and Friedman [5,6] calculated in detail the contribution of the inter-molecular potential for a bosonic system showing that it would need a repulsive potential, causing an expansion of the gas, in order to lower $T_B$ as one might expect from (7).

Shortly afterwards, ter Haar [7], pointed out that the repulsive potential was unphysical and would hinder the gas-liquid transition from taking place.

Recently, Deeney et al. [8] showed that a quantum source of energy leading to the expansion of the condensate may explain the negative $dT_\lambda/dP$ behavior. The QSHA model supports this hypothesis showing that the quantum pseudo potential (QPP) (that acts only in the quantum condensed state) generate a repulsive force leading to the anomalous behavior at lambda point.

The QPP is a well-defined potential energy in the Madelung's quantum hydrodynamic analogy (QHA). It is responsible for the realization of the eigenstates and the consequent quantum dynamics. As shown by Weiner [9], this energy is a real energy of the system and consists in the difference between the quantum energy and the classical one.

If fluctuations are present, the stochastic quantum hydrodynamic analogy (QSHA) shows that the quantum potential may have a finite range of interaction [10] so that dynamics owing a larger scale acquire the classical behavior. On the contrary on a scale shorter than the quantum coherence length $\lambda_c$ the quantum behavior is restored [10].

Following this approach, when the couples of $He^4$ molecules lie at a distance smaller or equal to the quantum coherence length $\lambda_c$, the atomic dynamics becomes quantum (the related quantum pseudo potential interaction appears) and the systems makes the $He^4_I \rightarrow He^4_{II}$ transition.

In the following the effects of the quantum pseudo potential energy onto the BEC temperature as well as on the sign of $dT_\lambda/dP$ are derived.

## 2. The QSHA equation of motion

The QHA-equations are based on the fact that the Schrödinger equation, applied to a wave function $\Psi_{(q,t)} = A_{(q,t)} \exp[i\, S_{(q,t)}/\hbar]$, is equivalent to the motion of a fluid with particle density $n_{(q,t)} = A^2_{(q,t)}$ and a velocity $\dot{q}_\alpha(q,t) = m^{-1} \partial S_{(q,t)}/\partial q_\alpha$, governed by the equations [11]

$$\partial_t n_{(q,t)} + \nabla_q \bullet (n_{(q,t)} \nabla_q \dot{q}) = 0, \tag{9.a}$$

$$\dot{q} = \nabla_p H, \tag{9.b}$$

$$\dot{p} = -\nabla_q (H + V_{qu}) \tag{9.c}$$

$$H = \frac{p \bullet p}{2m} + V_{(q)}. \tag{9.d}$$

By defining $\upsilon^H_j \equiv (\partial H/\partial p_\alpha, -\partial H/\partial q_\beta)$, $\upsilon^{qu}_j \equiv (0, -\partial V^{qu}/\partial q_\beta)$ we can ideally subdivide the phase-space velocity into the Hamiltonian and quantum part to read $\upsilon^Q_j = \upsilon^H_j + \upsilon^{qu}_j$. Moreover, $n$ is the number of structureless particles of the system whose mass is $m$ and $V^{qu}$ is the quantum pseudo-potential that originates the quantum non-local dynamics and reads

$$V_{qu} = -(\frac{\hbar^2}{2m}) n^{-1/2} \nabla_q \bullet \nabla_q n^{1/2}. \tag{10}$$

When fluctuations are considered into the hydrodynamic quantum equation of motion, the resulting stochastic QHA dynamics preserve the quantum behavior on a scale shorter than the theory defined quantum coherence length $\lambda_c$ [10]. Moreover, in the case of non-linear systems, on very large scale the local classical behavior can be achieved when the quantum pseudo potential has a finite range of interaction given by the non-locality length $\lambda_L$ [10] (with $\lambda_L > \lambda_c$).

Following the procedure given in reference [10], with $n_{(q,t)} = \int_{-\infty}^{+\infty} \rho_{(q,p,t)}\, dp_1 ... dp_{3n}$, where $\rho_{(q,p,t)}$ is the probability density function (PDF) of the system (whose spatial density $n_{(q,t)}$ represents the squared wave function modulus), the QSHA equation of motion can be established to read

$$\partial_t n_{(q,t)} = -\nabla_q \bullet (n_{(q,t)} \nabla_q \dot{q}) + \eta_{(q_\alpha, t, \Theta)}, \tag{11}$$

$$<\eta_{(q_\alpha,t)}, \eta_{(q_\alpha+\lambda,t+\tau)}> = \underline{\mu} \frac{4m(k\Theta)^2}{\pi^3 \hbar^2} exp[-(\frac{\lambda}{\lambda_c})^2]\delta(\tau)\delta_{\alpha\beta} \tag{12}$$

$$\lambda_c = (\frac{\pi}{2})^{3/2} \frac{\hbar}{(2mk\Theta)^{1/2}} \tag{13}$$

where $\Theta$ is a measure of the noise amplitude.

Moreover, given that (for the mono-dimensional case) the quantum potential range of interaction $\lambda_L$ (for $\lambda_L > \lambda_c$) reads [10]

$$\lambda_L = 2\lambda_c \frac{\int_0^\infty |q^{-1}\frac{dV_{qu}}{dq}|dq}{|\frac{dV_{qu}}{dq}|_{(q=\lambda_c)}}, \tag{14}$$

Where the origin (0,0) is the point of minimum Hamiltonian potential energy that is the rest mean position of the particle, for $\lambda_c \cup \lambda_L << \Delta\Omega$ equations (11-13) acquire the classical stochastic form

$$\partial_t n_{(q,t)} = -\nabla_q \bullet (n_{(q,t)} \nabla_q \dot{q}) + \eta_{(q_\alpha, t, \Theta)} \tag{15}$$

$$<\eta_{(q_\alpha,t)}, \eta_{(q_\alpha+\lambda, t+\tau)}> = \underline{\mu}\,\delta_{\alpha\beta}\frac{k\Theta}{\lambda_c}\delta(\lambda)\delta(\tau) \tag{16}$$

$$\dot{q} = \frac{p}{m} = \nabla_q \lim_{\Delta\Omega/\lambda_L \to \infty} \frac{S}{m} = -\nabla_q \{\lim_{\Delta\Omega/\lambda_L \to \infty} \frac{1}{m}\int_{t_0}^t dt(V_{(q)} + V_{qu}(n_0) + I^*)\}$$

$$= -\frac{1}{m}\nabla_q\{\int_{t_0}^t dt(V_{(q)} + I^*)\} = \frac{p_{cl}}{m} + \frac{\Delta p_{st}}{m} \tag{17}$$

## 3. Determination of the quantum potential at the $He^4_I \to He^4_{II}$ transition

In order to calculate the experimental outputs of the $He^4_I \to He^4_{II}$ superfluid transition we make use of the well-established statistical method of the Virial expansion that fits very fine for van der Waals fluids (see appendix [A]). This is possible in the stochastic hydrodynamic analogy since the presence of the quantum pseudo potential brings in the Virial expansion the quantum contribution to the system energy. A central point to derive the thermodynamic quantity by means of the Virial approach is the knowledge of the interaction in the pair of molecules (quantum potential included). Therefore, we firstly calculate the features of the $He^4 - He^4$ couple interaction.

As shown in ref. [19], the $He^4 - He^4$ interaction can be satisfying approximated by means of a square well potential of depth $\mathcal{U}^*$ and width $2\Delta$ such as

$$V_{LJ(q)} = \infty \quad x < \sigma \tag{18}$$
$$V_{LJ(q)} = -\mathcal{U}^* \qquad \sigma < x < \sigma + 2\Delta \tag{19}$$
$$V_{LJ(q)} = 0 \qquad x > \sigma + 2\Delta \tag{20}$$

(where $\sigma + \Delta$ is about the mean molecular (half) distance) and by introducing the self states wave functions

$$\psi = B \sin[K_n (x - \sigma)] \qquad\qquad \sigma < x < \sigma + 2\Delta \quad E_n > -\mathcal{U} \tag{21}$$

$$\psi = B \sin[K_n (2\Delta)] \exp[-\Gamma_n (x - (\sigma + 2\Delta))] \qquad x > \sigma + 2\Delta \quad E_n < 0 \tag{22}$$

where $\Gamma_n = (-2mE_n/\hbar^2)^{1/2}$, $K_n = (2m(\mathcal{U} + E_n)/\hbar^2)^{1/2}$, into relation (10), the quantum potential reads

$$V^{qu}{}_{(n)} = -(\hbar^2/2m)\,\Gamma_n{}^2 = E_n \qquad\qquad x > \sigma + 2\Delta \qquad (23)$$

$$V^{qu}{}_{(n)} = (\hbar^2/2m)\,K_n{}^2 = (\mathcal{U} + E_n) \qquad\qquad \sigma < x < \sigma + 2\Delta \qquad (24)$$

Where the values $E_n$ are given by the trigonometric equation

$$\tan[K_n(2\Delta)] = -K_n/\Gamma_n = -(-(\mathcal{U}+E_n)/E_n)^{1/2} \qquad E_n < 0 \qquad (25)$$

and hence

$$\Delta = (\hbar^2/8km)^{1/2}\,\arctan[-(-(\mathcal{U}+E_0)/E_0)^{1/2}] / (\mathcal{U}/k + E_0/k)^{1/2} =$$
$$= 1.231 \times 10^{-10} \{\arctan[-(-(\mathcal{U}+E_0)/E_0)^{1/2}]\} / (\mathcal{U}/k + E_0/k)^{1/2} \qquad (26)$$

Moreover, by assuming that the mean square well deepness $\mathcal{U}^*$ is slightly smaller than the L-J potential one $\mathcal{U}$ ($\mathcal{U}/k \cong 10.9\,°k$) and by evaluating that the value of the energy $E_0$ of the fundamental state at the transition is about

$$-E_0/k \sim T_{cr} = 5.19\,°K \qquad (27)$$

we obtain

$$\Delta \gtrsim 1.20 \times 10^{-10}\,m = 2.3\,\text{Bohr}. \qquad (28)$$

if we choose $\mathcal{U}^*$ to obtain the value for "a" given by (22) it follows that

$$\mathcal{U}^*/k \cong [V_{cr}/N_A((\sigma+2\Delta)^3 - (\sigma)^3)]\,\mathcal{U}/k = 0.82\,\mathcal{U}/k = 8.9\,°K,$$

from where, it follows that

$$\Delta \sim 1.54 \times 10^{-10}\,m = 2.9\,\text{Bohr} \qquad (29)$$

and that the mean $He^4{}_2$ atomic distance

$$\sigma + \Delta \cong 3.82 \times 10^{-10}\,m = 7.2\,\text{Bohr} \qquad (30)$$

that well agrees with the values 7.1 Bohr given in ref. [16]. Moreover, as shown in Appendix [B] the above results well agree with the Virial expansion applied to the $He^4$.

On this data, we can check that the coherence length of the deterministic quantum state $\lambda_c$ is coherently of order of the intermolecular distance at the $He^4{}_I \rightarrow He^4{}_{II}$ superfluid transition.

Reaching the lambda point (let's suppose by $He^4$ - $He^4$ cooling), the mean half atomic distance decreases to the value $\sigma + \Delta$ of the fundamental state and the wave function variance decreases to $2\Delta$ (the $He^4$ atoms lie almost inside the potential well). Therefore, assuming that the quantum coherence length $\lambda_c$ becomes much bigger that to the dimension of space domain where the wave function is relevant (i.e., the well width of $2\Delta$) to read

$$\lambda_c = \left(\frac{\pi}{2}\right)^{3/2} \frac{\hbar}{(2mk\Theta)^{1/2}} > 2\Delta, \qquad (31)$$

for the couple of $He^4$ - $He^4$ atoms, at lambda point it follows that

$$\Theta_\lambda < 0.59 \times 10^{-19}\,\Delta^{-2}\,°K \qquad (32)$$

and hence, by (28) that

$\Theta_\lambda < 4,0 \,°K$

or, more precisely, by using (29) that

$\Theta_\lambda < 2,49 \,°K$ (34)

Even if $\Theta$ is not exactly the thermodynamic temperature T, the result (47) is very satisfying since it correctly gives the order of magnitude of the transition temperature of the lambda point. The fact that $\Theta$ is close to T can be intuitively understood with the fact that going toward the absolute null temperature, correspondingly, $\Theta$ must decrease since the systems fluctuations must vanish in both cases.

As shown in [10] a relation between $\Theta$ and T can be established for an ideal gas at equilibrium. In this case, the thermodynamic temperature T converges to the vacuum fluctuation amplitude $\Theta$ in going toward the to absolute zero. In the case of a real gas and its fluid phase, a bit of difference between $\Theta$ and T may exists for $\Theta \neq 0$.

The result (34) definitely says that below a temperature of about 2,5°K degrees Kelvin the quantum potential enters more and more in the $He^4_I$–$He^4_I$ pair interaction. As it is shown in the following section, this well agrees with the features of the He lambda point that clearly shows how the increase of $He^4$ density (the sign of the quantum potential interaction) starts before the transition $He^4_I \rightarrow He^4_{II}$ takes place.

### 3.2. The sign of $\Delta T_\lambda = T_\lambda - T_B$ and of $dT_\lambda/dP$ at $He^4$ lambda point

The above equation (22) holds for normal fluid phases at a temperature above the superfluid transition one. Below the superfluid transition temperature, as shown by (31 and 47) the quantum coherence length $\lambda_c$ becomes larger than the inter-atomic $He^4$ - $He^4$ distance and hence the quantum potential contributes to the molecular energy and it must be taken into account in the calculation of the mean inter-molecular potential energy "a" that reads

$$a \equiv -2\pi \int_{r_0}^{\infty} (V_{LJ(r)} + V^{qu})\, r^2\, dr = -2\pi \left\{ \int_{r_0}^{\infty} V_{LJ(r)}\, r^2\, dr + \int_{r_0}^{\infty} V^{qu}\, r^2\, dr \right\} = a^{cl} + a^{qu} \quad (35)$$

where

$$a^{qu} = -2\pi \int_{r_0}^{\infty} V^{qu}\, r^2\, dr = -2\pi(\mathcal{U} + E_0) \int_{r_0}^{\sigma+\lambda_\mathcal{N}} r^2\, dr = -\tfrac{2}{3}\pi (\mathcal{U} + E_0)[(\sigma + \lambda_\mathcal{N})^3 - 2^{1/2}(\sigma)^3] < 0. \quad (36)$$

From (36) we can observe that $a^{qu}$ is negative since from (24) $V^{qu} = (\mathcal{U} + E_n)$ is positive. Therefore, below the superfluid transition temperature, the state equation (20) reads:

$$\{P + a^{cl}\, n^2/V^2 + a^{qu}\, n^2/V^2\}(V - nb) = \{P + a^{cl}\, n^2/V^2 + \Delta P^{qu}\}(V - nb) = n\, k\, T \quad (37)$$

where $\Delta P^{qu} = a^{qu}\, n^2 / V^2$, so that the pressure for $He^4_I$ and $He^4_{II}$ respectively reads

$$P_I(He^4_I) \cong \{n\, k\, T/(V - nb)\} - a^{cl}\, n^2/V^2 \quad (38)$$

$$P_{II}(He^4_{II}) = \{n\, k\, T/(V - nb)\} - a^{cl}\, n^2/V^2 - \Delta P^{qu} \quad (39)$$

Where it is posed $V \cong V_I \cong V_{II}$ since the fluid phase is poorly compressible. By using the same criterion of (7) the variation $\Delta T_\lambda = T_\lambda - T_{B \,(He^4_I)}$ has the same sign of the pressure difference to read

$$\Delta P = (P_I - P_{II}) = \Delta P^{qu} = a^{qu} n^2 / V^2 < 0 \qquad (40)$$

and the sign of $dT_\lambda /dP$ is the same of the derivative

$$d\Delta P^{qu} /dP = a^{qu} n^2 \, d\,(V^{-2})/dP < 0. \qquad (41)$$

given that $d(V^{-2})/dP$ is positive.

Therefore, the quantum potential of the QSHA leads to both $\Delta T_\lambda$ and $dT_\lambda /dP$ negative.

Finally, in order to show that the result obtained above is a direct consequence of the convex harmonic quantum potential, we use its more precise expression given in appendix [C], where the interaction of a couple of $He^4$ - $He^4$ atoms is approximated by a harmonic well (as given by the atomic Lennard-Jones potential) and coherently found to be

$$V^{qu}_{(q,t)} = -(\hbar^2/2m)|\psi|^{-1}\partial^2 |\psi|/\partial q \partial q = -(2\hbar^2/m) K_0^4 (q - \underline{q})^2 + (\hbar^2/m) K_0^2, \qquad (42)$$

where

where $K_0 = (2m(\mathcal{U} + E_0)/\hbar^2)^{½}$ and $\underline{q}$ is the mean $He^4$ - $He^4$ inter-atomic distance.

## 4. Discussion

The negative sign of both $\Delta T_\lambda = T_\lambda - T_B$ and of $dT_\lambda /dP$ are the direct consequence of the convex harmonic quantum potential that leads to a repulsive inter-atomic force so that the pressure of the superfluid $He^4_{II}$ is higher of that one it would assume the $He^4_I$ at the same temperature. Due to the repellent quantum potential energy, the passage from the $He^4_{II}$ state to the equivalent $He^4_I$ one (submitted to a lower pressure) needs less kinetic energy to happen and hence $T_\lambda$ is smaller than the condensation temperature $T_{B \,(He^4_I)}$. Moreover, since in $He^4_{II}$ a higher pressure than in $He^4_I$ is needed to maintain the same atomic distance, when the temperature is lowered at constant pressure near the lambda point (crossing $T_\lambda$) a decrease in density is produced as we get closer to the transition $He^4_I \rightarrow He^4_{II}$. Therefore, during the cooling process the $He^4$ shows a maximum in its density just above the $He^4_I \rightarrow He^4_{II}$ transition as confirmed by the experimental outputs.

It must be noted that for the realization of the maximum density, the crossover between the rate of change of the $He^4_I$ thermal shrinking and that one of the $He^4_{II}$ quantum dilatation is needed.

Moreover, since the density maximum is at 2.2 °K while $T_\lambda$ =2.17 °K, we can infer that the quantum interaction starts little bit before the transition temperature (i.e., $\Theta_\lambda < 2,49$ °K ) as (37) well signals (a larger and large fraction of $He^4$ atoms fall in the quantum interaction closer and closer we get to $T_\lambda$).

Moreover, it is worth mentioning that the QSHA model does not exclude the possibility of similar maximum density phenomena close to liquid-solid transitions (such as that of water) since, in this case, the quantum interaction between the atoms in a crystal is also set by the quantum potential whose interaction range $\lambda_L$ becomes larger than the typical inter-atomic distances [25]. This fact well agrees with the similarity between the $He^4_I \rightarrow He^4_{II}$ and the water-ice transitions widely accepted by the scientific community in the field. Also in this case, in order to have the maximum density at the liquid solid transition, the quantum dilation must overcomes the thermal shrinking velocity.

# 5. Conclusion

The finite range the quantum interaction in the QSHA is able to explain the controversial aspect of negative $dT_\lambda/dP$ at the $He^4$ lambda point without the introduction of a non-physical repulsive atomic potential that would hinder the gas-liquid phase transition [20,21]. The quantum pseudo-potential of the QSHA model is exactly the required potential: it is repulsive as widely requested by the scientific community to explain the maximum density of $He^4$ lambda point, but it also has the property to disappear in the classical phase and to not hinder the liquid-gas phase transition as any Hamiltonian potential would do.

The pseudo-potential of the QSHA approach also explains both why the lambda transition temperature $T_\lambda$ is smaller than the BE one $T_B$ and why the liquid $He^4$ has a maximum in its density just above the lambda point in agreement with the experimental measurements. The model puts in evidence that the perfect BE condensation is a phenomenon that happens between an ideal gas and its condensed quantum phase. As far as it concerns the liquid $He^4$, the phenomenon is slightly different being, by the fact, a transition between a real gas (in the fluid phase) and its quantum condensed phase so that the transition temperature is smaller.

Finally, it must be noted that even if the $He^4_I \rightarrow He^4_{II}$ is very well described by Montecarlo numerical simulation of standard quantum equations, the SQHA gives a modeling explanation that leads to a figurative comprehension of such a phenomenon that is complemental to that one coming by the numerical methods.

The SQHA kinetic equation is not a semi-empirical kinetic equation as those used to study the system behavior and its universality class near the phase transitions [22] but is a microscopic theoretical model from which exact Langevin kinetic equation can be obtained by the standard procedure of coarse-graining [23].

# Appendix A

## *The QSHA model for gas and condensed phases*

### Condensed phase and $He^4_I \rightarrow He^4_{II}$ transition

When both the lengths $\lambda_c$ and $\lambda_L$ are much smaller than the smallest physical length of the system (so that the resolution of the descriptive scale can be of order or bigger than $\lambda_L$) the macroscopic classical description arises. This for instance happens in a rarefied gas phase of L-J interacting particles where $\lambda_L$ as well as $\lambda_c$ are very small compared to the intermolecular mean distance (except for few colliding molecules).

On the contrary, when the mean inter-particle distance becomes comparable with the quantum non-locality length, the classical description may break down because the quantum potential enters in the particle interaction. Furthermore, if the wave function of the interacting particles is localized on a length of order or smaller than the quantum coherence length $\lambda_c$, the quantum deterministic description takes place for the bounded states of the couples of molecules.

In the classical regime, the Virial expansion furnishes an elegant conceptual understanding for passing from a gas to a condensed phase for molecules having finite range of interaction *even in non-equilibrium condition* [21].

In the classical treatment of the Virial expansion, the energy function does not include the quantum potential and hence converges to the classical value failing, for instance, to predict the law of the specific heat for solids where the quantum dynamics enters in the atoms interaction.

In the frame of the QHA description, the quantum potential energy (that changes at each stationary state) added to the classical value of the energy, leads to the variety of the quantum energy eigenvalues. This is very clearly shown in Ref. [9], the energy of the quantum eigenstates is composed by the sum of the two terms: one steams from the classical Hamiltonian while the other one by the quantum potential, leading to the correct eigenvalue $E_n$. Therefore, in principle the Virial approach can be applied (in the QSHA model) both for quantum as well classical molecular interactions

Since in a crystal the atoms fall in the linear range of interaction, the quantum non-locality $\lambda_L$ is larger than the inter-molecular distance (see Appendix [C]) and the system shows quantum characteristics (in those properties depending by the molecular state).

Usually, for crystalline solids the inter-atomic distance lies in the harmonic range of the L-J interaction even at temperature higher than the room one due to the great deepness of the potential well [25].

When, at higher thermal oscillations, the mean molecular distance starts to increases by the equilibrium position $r_0$ toward the non-linear range of the L-J inter-molecular potential, we have a transition from the solid phase to the liquid one [24]. During this process, the inter-particle wave function extends itself more and more in the non-linear L-J zone so that the quantum potential weakens and $\lambda_L$ decreases [25].

For deep L-J intermolecular potential well, this happens at high temperature and we have a direct transition from the solid to the classical fluid phase.

For small potential well, the liquid phase can persist down to a very low temperature. In this case, even if $\lambda_L$ may result smaller than the inter-particle distance (so that the liquid phase is maintained), decreasing the temperature, and hence the amplitude $\Theta$ of fluctuations, when $\lambda_c$ grows and becomes of order of the mean molecular distance, the liquid phase may acquire quantum properties (about those depending by the molecular interaction such as the viscosity). The fluid-superfluid transition can happen if the temperature of the fluid can be lowered up to the transition point before the solid phase takes place (i.e., $\lambda_L < r_0$).

Therefore, it worth noting that the mechanism that brings to the quantum inter-atomic interaction in a solid is different from that one in a superfluid: in the former the linearity of the interaction leads to a quantum non-locality length $\lambda_L$ larger than the typical atomic distance while in the latter is the decrease of $\Theta$, by lowering the temperature, that increases $\lambda_c$ up to the mean inter-atomic length.

Even if the relation between the PDF noise fluctuations amplitude $\Theta$ and the temperature T of an ensemble of particles is not straight [10], it can be easily acknowledged that when we cool a system toward the absolute zero (with steps of equilibrium) also the noise amplitude $\Theta$ reduces to zero since the energy fluctuations of the system must vanish. Thence, even there is not a fix linear relation between the fluctuation amplitude $\Theta$ and the temperature we expect lower values of $\Theta$ for lower values of the temperature [10].

# Appendix B

## *The Virial expansion applied to the He fluid*

As far as it concerns the first point, we have that from the standard Virial expansion [13] the state equation of (classical) real gas accounting only for double collisions, reads:

$$P V = n k T \{1 - (n (a/kT - b)/ V)\}, \qquad (18)$$

that under the standard substitution [14]

$$\{1 + n b / V\} \cong \{1 - n b / V\}^{-1} \qquad (19)$$

leads to the van der Waals equation

$$\{P + a n^2 /V^2 \} (V - n b) = n k T \qquad (20)$$

where P is the pressure, V the volume, n the number of molecules,

$$b = \tfrac{2}{3} \pi r_0^3 = V_{cr} / 3N_A \qquad (21)$$

is the fourfold atomic volume [15] and

$$a = -2 \pi \int_{r_0}^{\infty} V_{(r)} r^2 \, dr \cong 4\pi r_0^3 \mathcal{U}/3 = 2\mathcal{U} V_{cr} / 3N_A \qquad (22)$$

is the mean inter-molecular potential energy derived by using the rigid sphere approximation [13] that reads

$$V_{(r)} = \infty, \qquad x < r_0 \qquad (23)$$

$$V_{(r)} = V_{LJ(q)} = 4\mathcal{U} [(\sigma/q)^{12} - (\sigma/q)^6], \qquad x > r_0 \qquad (24)$$

where $\mathcal{U} = -V_{LJ(r_0)}$ is the well depth of the L-J intermolecular potential. Moreover, by using the relation [15]

$$a = 9 k T_{cr} V_{cr} / 8 N_A, \qquad (25)$$

from (22), for He$^4$ $\mathcal{U}/k$ reads

$$\mathcal{U}/k = 27\, T_{cr}/16 = 8.77°K, \tag{26}$$

satisfactory close to the value $\mathcal{U}/k = 10.9$ given by quantum Monte Carlo models [16] and to the value $\mathcal{U}/k = 11.07$ given by He$^4$-He$^4$ scattering [17].
Moreover, by using for helium [18] the value of

$$V_{cr} = 5.7 \times 10^{-5}\, m^3/moles, \tag{27}$$

it follows that

$$r_0 \cong 2.56 \times 10^{-10}\, m = 4.8\, bohr \tag{28}$$

where

$$r_0 = 2^{1/6}\sigma \tag{29}$$

is the point of minimum for the L-J intermolecular potential with

$$\sigma = 2^{-1/6}\, r_0 = 2.32 \times 10^{-10}\, m \cong 4.35\, Bohr.$$

# Appendix C

## *Quantum non-locality length of L-J bounded states*

In order to calculate the quantum potential and its non-locality length for a L-J potential well, we can assume the harmonic approximation

$$V_{LJ(q)} = \tfrac{1}{2}\, k\, (q - r_0)^2 + C, \tag{C.1}$$

where

$$k = 4\, K_0^4\, \hbar^2/m \tag{C.2}$$

where $K_0 = (2m(\mathcal{U} + E_0)/\hbar^2)^{1/2}$, and where the constant $C$ can be calculated by the energy eigenvalue of the fundamental state

$$C = E_0 - V^{qu}_{0\,(q-\underline{q}=0)}, \tag{C.3}$$

leading to a Gaussian wave function whose series expansion at second order coincides with that one of eqs. (21-22) having the same eigenvalue $E_0$ and mean position $\underline{q} = \sigma + \Delta$ that reads

$$\psi_0 = B\, \exp[-K_0^2(q - \underline{q})^2] \cong B\,[1 - K_0^2(q - \underline{q})^2] \cong B\, \sin[K_0(q - \sigma)] \qquad |q - r_0| << 2\Delta/\pi. \tag{C.4}$$

The convex quadratic quantum potential associated to the wave function $\psi_0$ reads

$$V^{qu}_{(q,t)} = -(\hbar^2/2m)|\psi|^{-1}\partial^2|\psi|/\partial q\partial q = -(2\hbar^2/m)\, K_0^4\,(q - \underline{q})^2 + (\hbar^2/m)\, K_0^2 \tag{C.5}$$

that leads to the quantum force

$$-\partial V^{qu}/\partial q = 2\, K_0^4\,(\hbar^2/m)\,(q - \underline{q}) \tag{C.6}$$

and to

$$C = E_0 - V^{qu}{}_0{}_{(q=0)} = \tfrac{1}{2}\hbar(k/m)^{1/2} - (\hbar^2/m) K_0^2 = 0.$$

Given the simple exponential PDF decrease of (21-22) for x > σ + 2Δ (that leads to a vanishing quantum potential as well as to vanishing small quantum force), we can disregard the contribution to the quantum non-locality length for x > σ + 2Δ.
Thence, by (13) it follows that,

$$\lambda_L \cong 2\lambda_c \frac{\int_{\sigma+\Delta}^{\sigma+2\Delta} |q^{-1}\frac{dV_{qu}}{dq}|\,dq}{|\frac{dV_{qu}}{dq}|_{(q=\lambda_c)}} \tag{C.7}$$

$|d V^{qu}/dq| = 2 K_0^4 (\hbar^2/m) \lambda_c$

$|-q^{-1}\partial V^{qu}/\partial q| = |2 K_0^4 (\hbar^2/m)|$

$$\lambda_L \cong 2 \int_{\sigma+\Delta}^{\sigma+2\Delta} dq = 2\Delta = \lambda_c$$

# Appendix D

## *Pseudo-Gaussian PDF*

If a system admits the large-scale classical dynamics, the PDF cannot acquire an exact Gaussian shape because it would bring to an infinite quantum non-locality length.
In section (III.B.1) we have shown that for $h < 3/2$ (when the PDF decreases slower than a Gaussian) a finite quantum length is possible.
The Gaussian shape is a physically good description of particle localization but irrelevant deviations from it, at large distance, are decisive to determine the quantum non-locality length.
For instance, let's consider the pseudo-Gaussian function type

$$n_{(q,t)} = \exp[-(q-\underline{q})^2/<\Delta q^2>][1 + [(q-\underline{q})^2/\Lambda^2 f(q-\underline{q})]]], \tag{D.1}$$

where $f(q-\underline{q})$ is an opportune regular function obeying to the conditions

$\Lambda^2 f(0) >> <\Delta q^2>$ and $\lim_{|q-\underline{q}|\to\infty} f(q-\underline{q}) << (q-\underline{q})^2 / \Lambda^2$.

For small distance $(q-\underline{q})^2 << \Lambda^2 f(0)$ the above PDF is physically indistinguishable from a Gaussian, while for large distance we obtain the behavior

$$\lim_{(q-\underline{q})\to\infty} n_{(q,t)} = \exp[-\Lambda^2 f(q-\underline{q})/<\Delta q^2>]. \tag{B.3}$$

For instance, we may consider the following examples

i.     $f(q-\underline{q}) = 1$         $\lim_{|q-q_0|\to\infty} n_{(q,t)} = \exp[-\Lambda^2/<\Delta q^2>]$ ;         (D.4)

|     |                                                                 |                                                                                           |       |
| --- | --------------------------------------------------------------- | ----------------------------------------------------------------------------------------- | ----- |
| ii. | $f(q-\underline{q}) = 1 + |q-\underline{q}|$                    | $\lim_{|q-q_0|\to\infty} n_{(q,t)} \propto \exp[-\Lambda^2|q-\underline{q}|/<\Delta q^2>]$ ; | (D.5) |
| iii.| $f(q-\underline{q}) = 1 + \ln[1+|q-\underline{q}|^h]$ $(0< h <2)$ | $\lim_{|q-q_0|\to\infty} n_{(q,t)} \propto |q-\underline{q}|^{-h\,\Lambda^2/<\Delta q^2>}$ ; | (D.6) |
| iv. | $f(q-\underline{q}) = 1 + |q-\underline{q}|^h$ $(0< h <2)$       | $\lim_{|q-q_0|\to\infty} n_{(q,t)} \propto \exp[-\Lambda^2|q-\underline{q}|^h/<\Delta q^2>]$. | (D.7) |

All cases (i-iv) lead to a finite quantum non-locality length $\lambda_L$.

In the case (iv)(D.1) reads

$$n_{pg}(q,t) = \exp[-((q-\underline{q})^2/<\Delta q^2>\{1 + \Lambda^{-2}(q-\underline{q})^2/|q-\underline{q}|^h\})] \quad (h<2). \tag{D.8}$$

Given that for the PDF(D.8)

$$\lim_{|q-\underline{q}|\to\infty} |\psi| = \lim_{|q-\underline{q}|\to\infty} n_{pg}^{1/2} = \exp[-\Lambda^2(q-\underline{q})^h/2<\Delta q^2>],$$

the quantum potential for $|q| >> |\underline{q}|$ reads:

$$\lim_{(q-\underline{q})\to\infty} V^{qu} = -(\hbar^2/2m)|\psi|^{-1}\partial^2|\psi|/\partial q \partial q = -(\hbar^2/2m)[\Lambda^4 h^2(q-\underline{q})^{2(h-1)}/4<\Delta q^2>^2) - h(h-1)(q-\underline{q})^{(h-2)}], \tag{D.9}$$

leading, for $h \neq 2$, to the quantum force

$$\lim_{(q-q_0)\to\infty} \partial V^{qu}/\partial q = -(\hbar^2/2m)[\Lambda^4(2h-1)h^2(q-\underline{q})^{2h-3}/4<\Delta q^2>^2) - \Lambda^2 h(h-1)(h-2)(q-\underline{q})^{(h-3)}/2<\Delta q^2>], \tag{D.10}$$

that for $h < 3/2$ gives $\lim_{(q-\underline{q})\to\infty} \partial V^{qu}/\partial q = 0$.